# Generation of phase-matched circularly-polarized extreme ultraviolet high harmonics for magnetic circular dichroism spectroscopy

Ofer Kfir[1], Patrik Grychtol[2], Emrah Turgut[2], Ronny Knut[2,3], Dmitriy Zusin[2], Dimitar Popmintchev[2], Tenio Popmintchev[2], Hans Nembach[2,3], Justin M. Shaw[3], Avner Fleischer[1,4], Henry Kapteyn[2], Margaret Murnane[2] and Oren Cohen[1].

[1]*Solid State Institute and Physics Department, Technion, Haifa 32000, Israel*
[2]*Department of Physics and JILA, University of Colorado and NIST, Boulder, CO 80309, USA*
[3]*Electromagnetics Division, National Institute of Standards and Technology, Boulder, CO 80305, USA*
[4]*Department of Physics and Optical Engineering, Ort Braude College, Karmiel 21982, Israel*
Corresponding authors: *ofertx@technion.ac.il*, *oren@technion.ac.il*

Circularly-polarized extreme UV and X-ray radiation provides valuable access to the structural, electronic and magnetic properties of materials. To date, this capability was available only at large-scale X-ray facilities such as synchrotrons. Here we demonstrate the first bright, phase-matched, extreme UV circularly-polarized high harmonics and use this new light source for magnetic circular dichroism measurements at the *M*-shell absorption edges of Co. We show that phase matching of circularly-polarized harmonics is unique and robust, producing a photon flux comparable to the linearly polarized high harmonic sources that have been used very successfully for ultrafast element-selective magneto-optic experiments. This work thus represents a critical advance that makes possible element-specific imaging and spectroscopy of multiple elements simultaneously in magnetic and other chiral media with very high spatial and temporal resolution, using tabletop-scale setups.

**Introduction**

Circularly polarized radiation in the extreme ultraviolet (EUV) and soft X-ray spectral regions has proven to be extremely useful for investigating chirality-sensitive light-matter interactions. It enables studies of chiral molecules using photoelectron circular dichroism[1], ultrafast molecular decay dynamics[2], the direct measurement of quantum phases (*e.g.* Berry's phase and pseudo-spin) in graphene and topological insulators[3–5] and reconstruction of band structure and modal phases in solids[6]. For magnetic materials, circularly polarized soft x-rays are particularly useful for X-ray Magnetic Circular Dichroism (XMCD) spectroscopy[7]. XMCD enables element-selective probing as well as coherent imaging and holography of magnetic structures with nanometer resolution[8–10]. Moreover, it can also be used to extract detailed information about the magnetic state by distinguishing between the spin and orbital magnetic moments of each element. Thus, time-resolved XMCD can probe the element-specific dynamics of the spin and orbital moments when interacting with the electronic and phononic degrees of freedom in a material[11–14]. However, the time resolution available to date for XMCD has been > 100 fs, limited by the pulse duration and timing jitter of synchrotron pulses[15–17]. To date it has not been possible to probe spin dynamics of multiple elements simultaneously within the same sample, because the photon energy must be tuned across the various absorption edges at the large-scale facilities where these experiments are currently performed.

Table-top soft x-ray sources based on high harmonic upconversion of femtosecond laser pulses represent a viable alternative to large-scale sources for many applications, due to their unique ability to generate bright, broadband, ultrashort and coherent light with an energy spectrum reaching into the keV region[18]. High harmonic generation (HHG) not only enables coherent imaging of nanometer structures with a spatial resolution approaching the diffraction limit[19], but also accesses the fastest dynamics in atoms, molecules, solids and plasmas with unprecedented time resolution[20–27]. The large bandwidth and short temporal duration of HHG makes it a versatile, element-specific probe of coupled spin, charge and phonon dynamics on ultrafast timescales. In the case of magnetic materials, recent work has shown that HHG can simultaneously probe the magnetic state of the 3d ferromagnets Fe, Co and Ni, by taking advantage of the fact that the reflectivity near the *M*-shell absorption edges depends on the orientation as well as the magnitude of the magnetization[28,29]. This novel capability made it possible to uncover new fundamental understanding about the timescale of the exchange interaction, as well as revealing the important roles of local spin scattering and non-local spin transport – all of which can occur on few-femtosecond timescales[30–32]. The vast majority of HHG applications to date, however, used linearly polarized light that can be generated relatively efficiently.

For decades it was assumed that high harmonic generation from atoms was brightest when both the driving laser and HHG fields were linearly polarized. The source for this misconception is that HHG is a recollision phenomenon. A bound electron is ripped from an atom by the intense electric field of a driving laser pulse and then accelerated as a free electron until it re-encounters its parent ion[33,34]. For a linearly polarized driving field, the electron accelerates on a linear trajectory - and therefore easily recollides with the parent ion. When driven by a slightly elliptical laser field, the electron has some probability to recollide with its parent ion due to lateral spreading (quantum diffusion) of the wave function in the continuum[35,36]. This results in the generation of slightly elliptically polarized high harmonics[37,38]. In contrast, for circularly-polarized driving lasers (or elliptically-polarized laser with a large ellipticity), the probability for recollision and emission of high harmonics is completely suppressed. Nevertheless, because of important applications in materials science, there was strong motivation to generate bright circularly polarized high harmonic beams[39–49]. The first experimentally-measured circularly polarized high-order harmonics were produced by converting the polarization of linearly polarized HHG to circular, using a reflective quarter-waveplate[45]. This approach is very lossy, is limited to narrow spectral regions constrained by available multilayer mirror materials, and to date has not proven useful for applications. A direct approach for generating circularly-polarized HHG was suggested almost two decades ago[50,51] and recently measured by Fleischer *et al.*[52]. In this scheme, circularly polarized high harmonics are driven by co-propagating circularly-polarized bi-chromatic fields that rotate in opposite directions and interact with an isotropic gas. Notably, Fleischer *et al.* measured the ellipticity of weak harmonics generated from HHG in a thin gas jet, but not their helicity (i.e. right vs. left-circular polarization).

In this work, we generate bright, phase-matched, circularly-polarized high harmonic beams for the first time, and then use this unique tabletop light source to implement the first magnetic circular dichroism measurements of magnetic materials. To achieve this, we first demonstrate both experimentally and theoretically that phase matching of high harmonics driven by circularly polarized bi-chromatic waves is significantly different than all other approaches to phase matching to date. Phase matching is more robust for circularly-polarized bi-chromatic drivers than for HHG with bi-chromatic linearly- or elliptically-polarized drivers. Moreover, phase matching can selectively enhance either left or right circularly polarized harmonics. Using this new understanding, we generate bright circularly-polarized high harmonics by co-propagating bi-chromatic driving laser beams (the fundamental and second harmonic beams from a Ti:Sapphire laser) that are circularly-polarized with opposite helicity in a gas-filled hollow waveguide. (Section 6 in the supplementary information (SI) elaborates on the important features of this source). We show that bright, circularly-polarized, high harmonics in Ne can span multiple *M*-edges of the 3d ferromagnets, and obtain a photon flux of ~$10^9$ HHG photons/s per harmonic (section 5 in SI), which is comparable to the linearly-polarized HHG flux that was used very successfully in previous ultrafast element-selective magneto-optic HHG-based experiments[28–32]. Finally, we measure the circular dichroism of Co throughout the *M*-shell absorption edge spectral region. This measurement is also the first to measure the helicity of circularly polarized high-order harmonic beams. In the future, the use of mid-infrared pump lasers should make it possible to generate bright circularly-polarized high-harmonics at the L-shell absorption edges of Fe, Co, and Ni, in the photon energy range approaching 1 keV[18].

**Generation of and phase matching condition for circularly-polarized HHG**

We first describe the theoretical spectral and polarization features of our source driven by a left-rotating circularly-polarized fundamental beam at a central wavelength of 790 nm, and a right-rotating circularly-polarized second harmonic beam (Fig. 1a). We consider an idealized case in which the medium is isotropic and time-independent and where the driving laser is perfectly periodic. The bi-chromatic field, $\vec{E}_{BC}(t)$, exhibits the following dynamical symmetry, independently of the relative intensity or phase between the two pump fields - $\vec{E}_{BC}(t + T/3) = \hat{R}_{(120°)}\vec{E}_{BC}(t)$. $T$ is the optical cycle of the fundamental laser and $\hat{R}_{(120°)}$ is the 120° rotation operator in the polarization plane. The emitted HHG field conforms to the same dynamical symmetry. Thus, as can be verified by taking the Fourier transform of the dynamical symmetry constraint, the complex amplitude of each $q^{th}$-order harmonic, $\vec{E}_q$, satisfies the following eigenvalue equation

$$\vec{E}_q e^{-2\pi i q/3} = \hat{R}_{(120°)}\vec{E}_q \qquad (1)$$

The solutions of Eq. (1) are a left circularly-polarized field with eigenvalue $e^{-2\pi i/3}$ and a right circularly-polarized field with eigenvalue $e^{+2\pi i/3}$. The left circularly-polarized $q^{th}$-order harmonic, $\vec{E}_{q,L}$, satisfies $e^{-(2\pi i/3 \cdot q)}\vec{E}_{q,L} = e^{-(2\pi i/3)}\vec{E}_{q,L}$ which is fulfilled only when $q=3m+1$. Similarly, the right circularly-polarized $q^{th}$-order harmonic, $\vec{E}_{q,R}$, satisfies $e^{-(2\pi i/3 \cdot q)}\vec{E}_{q,R} = e^{+(2\pi i/3)}\vec{E}_{q,R}$ which is fulfilled only when $q=3m-1$. Notably, $q=3m$ harmonics do not satisfy Eq. (1) and are therefore theoretically forbidden.

Phase matching is essential for obtaining a bright and collimated high harmonic beam. In HHG with linearly-polarized bi-chromatic drivers, the phases and intensities of the emitted harmonics depend strongly on the relative phase between the chromatic drivers, due to the intrinsic phases of the rescattering electrons[53]. Consequentially, full phase matching in this case requires that the two driver fields and the harmonic field propagate at the same phase velocity; a very challenging condition to satisfy. Fortunately, phase matching of HHG driven by circularly-polarized bi-chromatic drivers is very different. In this case, the relative phase between the chromatic driver components does not influence the shape (cloverleaf) of the bi-chromatic driver, and hence it also does not influence the path length of the recolliding electron and the intrinsic phase it accumulates. Thus, phase mismatch of the process is determined only by the extrinsic phase, and is described by the 'ordinary' phase-mismatch equation in nonlinear optics:

$$\frac{\pi}{L_c} = \Delta k = k_q - \ell k_1 - m k_2 \qquad (2)$$

In Eq. (2) $k_q, k_1, k_2$ are the wave vectors for the $q^{th}$ order high-harmonic, fundamental and second harmonic beams, respectively, $\Delta k$, is the phase mismatch of the upconversion process, and $L_c$ is the resulting coherence length. $\ell$ and $m$ are the number of photons of the fundamental and second harmonic pump beams, respectively, converted to a $q^{th}$ order high-harmonic photon. Conservation of energy and angular momentum impose that $q = \ell + 2m$, and $\ell = m \pm 1$, respectively[52].

For simplicity, we associate $k_1$ and $k_2$ with the effective index change, $\Delta n_1$ and $\Delta n_2$, respectively, defined as $\Delta n_i = (k_i/[2\pi/\lambda_i] - 1)$, where $\lambda_i$ is the vacuum wavelength of the $i^{th}$ driver ($i=1$ for the fundamental pump and $i=2$ for the second harmonic pump). Thus, we can rewrite Eq. (2) in the form:

$$\frac{\pi}{L_c} = -\frac{2\pi}{\lambda_1}[m(\Delta n_1 + 2\Delta n_2) \pm \Delta n_1] \qquad (3)$$

where the + and − signs correspond to the $q=3m+1$ and $q=3m-1$ groups of harmonics, respectively. The factor $\pm \Delta n_1$ leads to a large difference in the coherence lengths between the combs of left and right rotating circular harmonics. Figure 1b shows numerical results for phase matching of circularly polarized HHG. The coherence lengths shown in Fig. 1b are calculated for $\Delta n_1$ and $\Delta n_2$ that optimize phase matching for the $28^{th}$ harmonic. As expected, the q=3m+1 harmonics are brighter (i.e. better phase matched to the bi-chromatic pump) than the harmonics with opposite helicity. Notably, in order to implement phase matched HHG frequency upconversion, the ionization level of the gas must be below a critical value, $\eta_{cr}$[54]. For HHG with a circularly polarized bi-chromatic driver, $\eta_{cr}$ has a value between the critical ionization for HHG driven by the fundamental or second harmonic, separately. (See Eq. (S.3) in the supplementary information). This also results in a phase matching cutoff photon energy that is somewhere between the cutoff corresponding to the fundamental and the second harmonic, separately.

Figure 2 presents a schematic of the experimental setup. We use a Ti:Sapphire oscillator in combination with a single-stage regenerative amplifier, delivering sub-45 fs pulses at a repetition rate of 4 kHz, which are centered at a wavelength of 790 nm (red), with an energy of 2.5 mJ per pulse. After frequency doubling the laser beam in a beta-phase barium borate (BBO) crystal, the bi-chromatic co-propagating driving fields are separated by a dichroic mirror (DM) into two different arms of a Mach-Zehnder interferometer. An optical delay stage in the red arm compensates for the relative time delay between the two colors, and the polarization of each arm is fully controlled by a pair of half- and quarter-wavelength retardation plates. The red (1.6 mJ/pulse) and blue (0.43 mJ/pulse) driving lasers are focused into a 150 μm diameter, 2 cm long gas-filled hollow waveguide using lenses with 50 cm and 75 cm focal lengths, respectively. The circularly polarized HHG beam that emerges from the waveguide then passes through an aluminum filter to block the pump laser beams. In our first experiment, the HHG beam is then spectrally dispersed using a spectrometer composed of a toroidal mirror, a laminar gold grating with a groove density of 500 lines/mm, and a CCD camera. In our second experiment, the HHG beam passes through a magnetized 100 nm thick Co foil tilted by 45° with respect to the harmonics beam before entering the spectrometer. The magnetization of the Co foil is controlled by an external electromagnet that produces a magnetic field of approximately $4.8 \cdot 10^4$ A/m (60 mT) parallel to the sample plane, exceeding the $1.6 \cdot 10^4$ A/m (20 mT) coercive field of the foil. The alternating polarity of the magnet is synchronized to the spectrometer read-out, so that we can acquire the transmitted HHG spectrum for two ("up" and "down") magnetization directions. The normalized difference of these two spectra constitutes the XMCD asymmetry. Positioning the ferromagnetic sample at 45° degrees to the HHG beam allows observation of XMCD of a sample that is magnetized in plane[55].

We first discuss the brightness, spectrum, and polarization of the EUV light produced by our novel HHG source. Figure 3a shows the observed HHG spectrum when the hollow waveguide was filled with Ar gas at a pressure of 100 Torr (13 kPa), where the driving laser beams were blocked by Al filters of ~800 nm thickness. Figure 3b displays the observed HHG spectra from 650 Torr (87 kPa) of Ne, and 70 Torr (9 kPa) of $N_2$ using 400 nm thick aluminum filters. Clearly, the *q=3m* harmonics are almost completely suppressed over the entire observed HHG spectra for all gas species. This feature indicates that the polarization of the harmonics is near circular[52].

Figures 3c and 3d present, respectively, an experimental and numerical pressure-dependent spectrogram of HHG emission from a Ne-filled hollow waveguide (see section 3 in the supplementary information for details). Two features are worth mentioning: First, the 3m HHG orders are suppressed throughout the pressure range, reflecting the single atom dynamical symmetry that was described in Eq. (1). Second, the intensity of the plateau harmonics dramatically increases at pressures above 100 Torr (13 kPa), indicating that they are phase matched throughout this pressure range. However, the cut-off harmonics are phase matched within a narrower pressure range, marked by yellow dashed lines. This effect, which also exists in HHG from linearly polarized driving fields[54], results from two complementary effects: First, a larger effective index change $|\Delta n_1|, |\Delta n_2|$, is required for phase matching higher order HHG. Thus, deviation from perfect phase matching in Eq. (2), say 1% of $\Delta n_{1,2}$, affects the absolute value of the coherence length $L_c$ more drastically. Second, the longer re-absorption length of higher HHG orders in Ne[56] makes the phase-matched throughput even more sensitive to imperfect phase matching conditions. Notably, re-absorption of the high harmonics in Ne limits the distinction between full and partial phase matching. Therefore, the output flux contrast between one perfectly phase matched harmonic (e.g. 28$^{th}$ order in Fig. 1b) and the rest of the 3$m$+1 harmonic orders is insignificant. Most importantly, the current flux of circularly-polarized photons produced by our scheme is sufficient for application experiments such as XMCD, as described next.

**Magnetic circular dichroism**

To demonstrate the utility of this table-top HHG source with circularly-polarized EUV light for probing magnetization, we measured XMCD of a freestanding thin Co foil. The imaginary, *i.e.* absorptive, part of the refractive index, $n$, for light propagating in a magnetically saturated film is given by $Im[n] = \beta \pm \Delta\beta$ where $\beta$ is the average absorption coefficient, and $\Delta\beta$ is the dichroic absorption coefficient for circularly polarized light, where the ($\pm$) marks the absorption difference for left versus right-circularly polarized EUV photons[55,57]. The dichroic absorption scales linearly with the component of the magnetization parallel to the direction of light propagation. For example, the absorption coefficient of a sample with magnetization $M$ that is below its saturation value, $M_{sat}$, pointing at angle $\theta$ with respect to the beam direction is $(M/M_{sat})\cos\theta \cdot \Delta\beta$. However, XMCD requires a bright circularly polarized soft X-ray source, since the non-dichroic absorption coefficient is $\approx$ 10 times larger than the dichroic coefficient, leading to strong average absorption that far exceeds the dichroic signal.

We measured the HHG spectrum after passing through a magnetized Co foil, as shown in Figs. 2 and 4. Figure 4a presents the transmitted spectrum – $I^{up}$ and $I^{down}$ for Co magnetization of "up" and "down", respectively. The normalized XMCD asymmetry, $A = (I^{up} - I^{down})/(I^{up} + I^{down})$ is shown in Fig. 4b. The asymmetry of the 3$m$+1 and 3$m$-1 harmonics exhibit opposite signs - proving that the helicity of the 3$m$+1 harmonics is indeed opposite to the 3$m$-1 harmonics.

Finally, we used the measured XMCD asymmetry to extract the magneto-optical (MO) dichroic absorption coefficient, $\Delta\beta$, for Co. The MO dichroic absorption coefficient is given by $\Delta\beta = \tanh^{-1}(A)/(2|\vec{k}_0|\cos\theta\, L)$, where $\vec{k}_0$ is the light wave vector, $L$ is the optical path length in the sample and $\theta$ is the angle between the sample magnetization and beam propagation direction[55]. Since the sample magnetization is tilted by 45°, the optical path is $\sqrt{2}$ times the sample thickness, $d$, and $\cos\theta = 1/\sqrt{2}$, the MO dichroic coefficient is simply $\Delta\beta = \tanh^{-1}(A)/(2|\vec{k}_0|d)$. Our results for $\Delta\beta$ are presented in Fig. 4c. For each HHG order, we estimate $\Delta\beta$ from a weighted average over a harmonic peak, where the error bars are the standard deviation over the same harmonic. The MO coefficient measured by the left circularly polarized 3$m$+1 harmonics matches measurements that were taken using synchrotrons[57], indicating that their polarization is circular. However, the low intensity and low MO dichroic absorption coefficient measured using the right circularly polarized harmonics 29$^{th}$, 32$^{nd}$ and 35$^{th}$ (corresponding to the 3$m$-1 group) at 45 eV, 50 eV and 55 eV, respectively, indicate that their polarization is elliptical and not fully circular. Such imperfections may arise since the comb of forbidden harmonics ($q$=3$m$) are not perfectly suppressed. The reasons for the reduced circularity of the 3*m*-1 as opposed to the 3*m*+1 harmonics are not yet clear and are a subject of continued investigation.

**Conclusions and outlook**

We demonstrated the first bright, phase matched, source of circularly polarized high harmonics and also demonstrated the first *M*-edge X-ray magnetic circular dichroism measurements of magnetic materials on a tabletop. Our HHG source produces a broad spectrum of bright circularly-polarized harmonics, where consecutive harmonics exhibit opposite helicity, and where phase matching consideration are robust and favor one helicity over the other. This work removes a major constraint to date - that the polarization of bright high harmonics sources was limited to linear polarization - and therefore paves the way for ultrafast circular dichroism studies of magnetism, chiral molecules and nanostructures. In combination with coherent diffraction imaging techniques, this advance will also enable the exploration of coupled spin, charge and structural dynamics of magnetic domains with elemental specificity and unprecedented spatial and temporal resolution. Our source might also be used for seeding x-ray free electron lasers to obtain ultra-bright circularly-polarized X-rays[58]. Finally, the scheme we use is universal, and can be used to generate circularly-polarized high harmonics across broad spectral regions, for example, through quasi phase matching of the HHG process, by using mid-IR pumps[18] or by using other nonlinear media.

**Acknowledgements**

This work was supported by the USA–Israel Binational Science Foundation (BSF). The Technion group is part of the Israeli Center of Research Excellence `Circle of Light' supported by the I-CORE Program of the Planning and Budgeting Committee and The Israel Science Foundation. The JILA group gratefully acknowledges funding from the U.S. Department of Energy Office of Basic Energy Sciences, Award #DE-SC0002002**,** from the Deutsche Forschungsgemeinschaft #GR 4234/1-1, and from the Physics Frontiers Center Program. JILA also gratefully acknowledges support from an AFOSR DURIP award for the laser system used for this work.

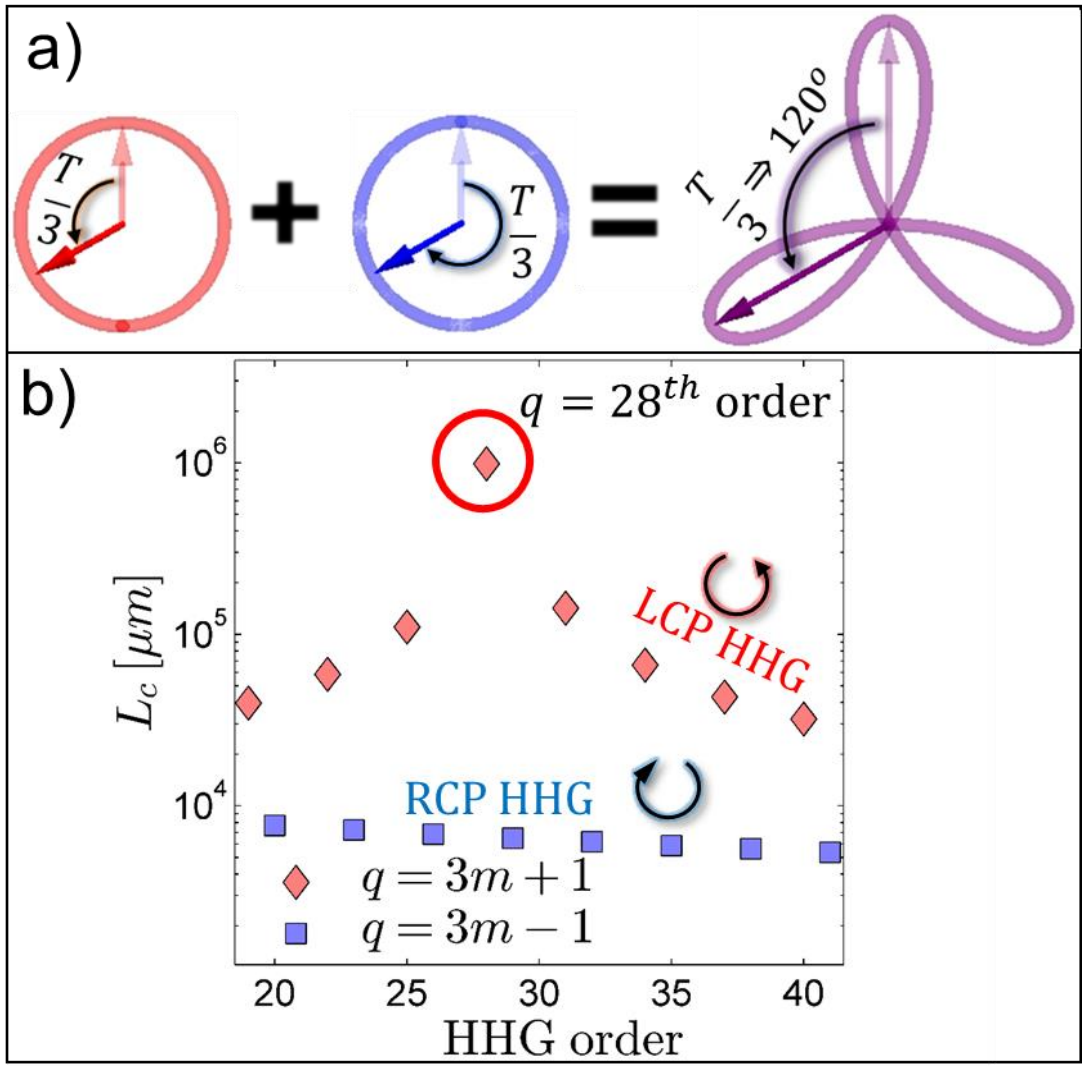

**Figure 1** | (a) The combined electric field of a circularly polarized 790 nm driving laser beam (red) and a counter rotating second harmonic field (blue) is a three-fold cloverleaf shape (purple). The system possess a discrete rotation dynamical symmetry of the light field, where a delay of $T/3$ acts as a $120^o$ rotation. This dynamical symmetry gives rise to circularly polarized harmonics where harmonic orders $q=3m-1$ rotate right, while $q=3m+1$ rotate left. (b) Calculated coherence length versus HHG order under phase matching conditions that optimize the 28$^{th}$ harmonic. Phase matching of circularly polarized HHG is helicity selective, so HHG orders $q=3m+1$ (left-circular polarization (LCP)) have a longer coherence length (better phase matching) than $q=3m-1$ (right-circular polarization (RCP)). Re-absorption of high harmonics in the gas medium reduces the contrast arising from this effect, as elaborated on the supplementary information.

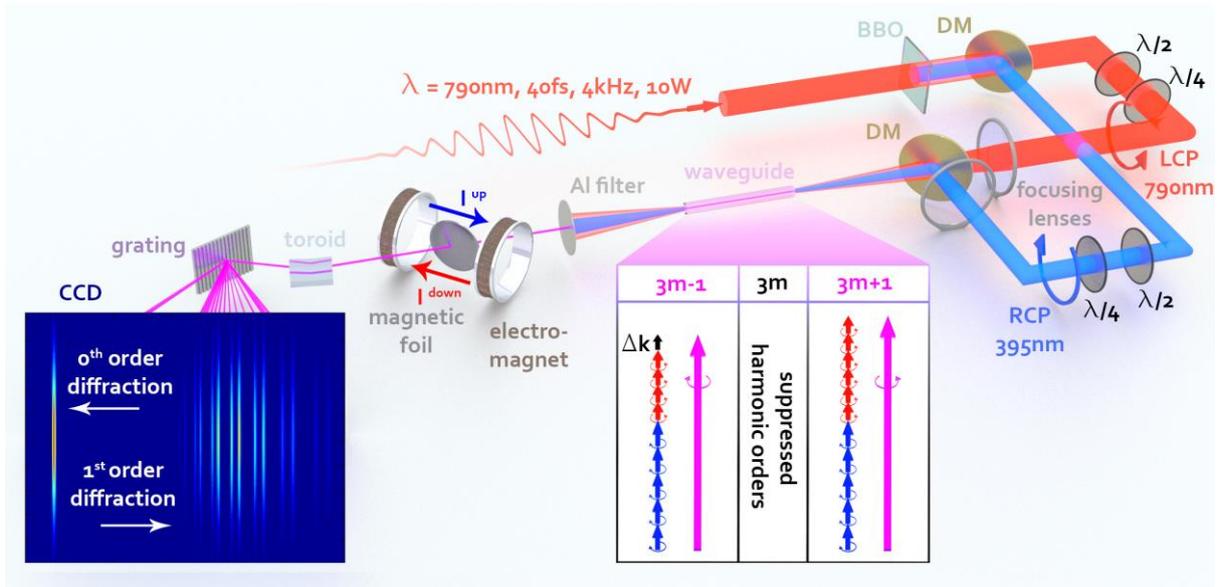

**Figure 2 | Experimental apparatus.** Circularly-polarized harmonics are generated by focusing fundamental and second harmonic pump beams (with opposing helicities) into a gas filled hollow waveguide. The polarization states of the two drivers can be independently adjusted, thereby controlling the helicity of the generated harmonics. The magnetic state of a sample can then be measured with elemental-specificity via XMCD by use of harmonics that overlap with an appropriate core-level absorption edge, e.g., the *M*-edge of Co. The right inset Illustrates the helicity selectiveness for phase matching of circularly polarized HHG, where left rotating harmonics ($q=3m+1$) and right rotating harmonics ($q=3m-1$) cannot be phase matched simultaneously. I.e. if the $q=3m+1$ harmonics are phase matched with the pump lasers, the $q=3m$-1 are mismatched by $\Delta k$. The left inset illustrates a typical experimental spectrum as recorded on the CCD.

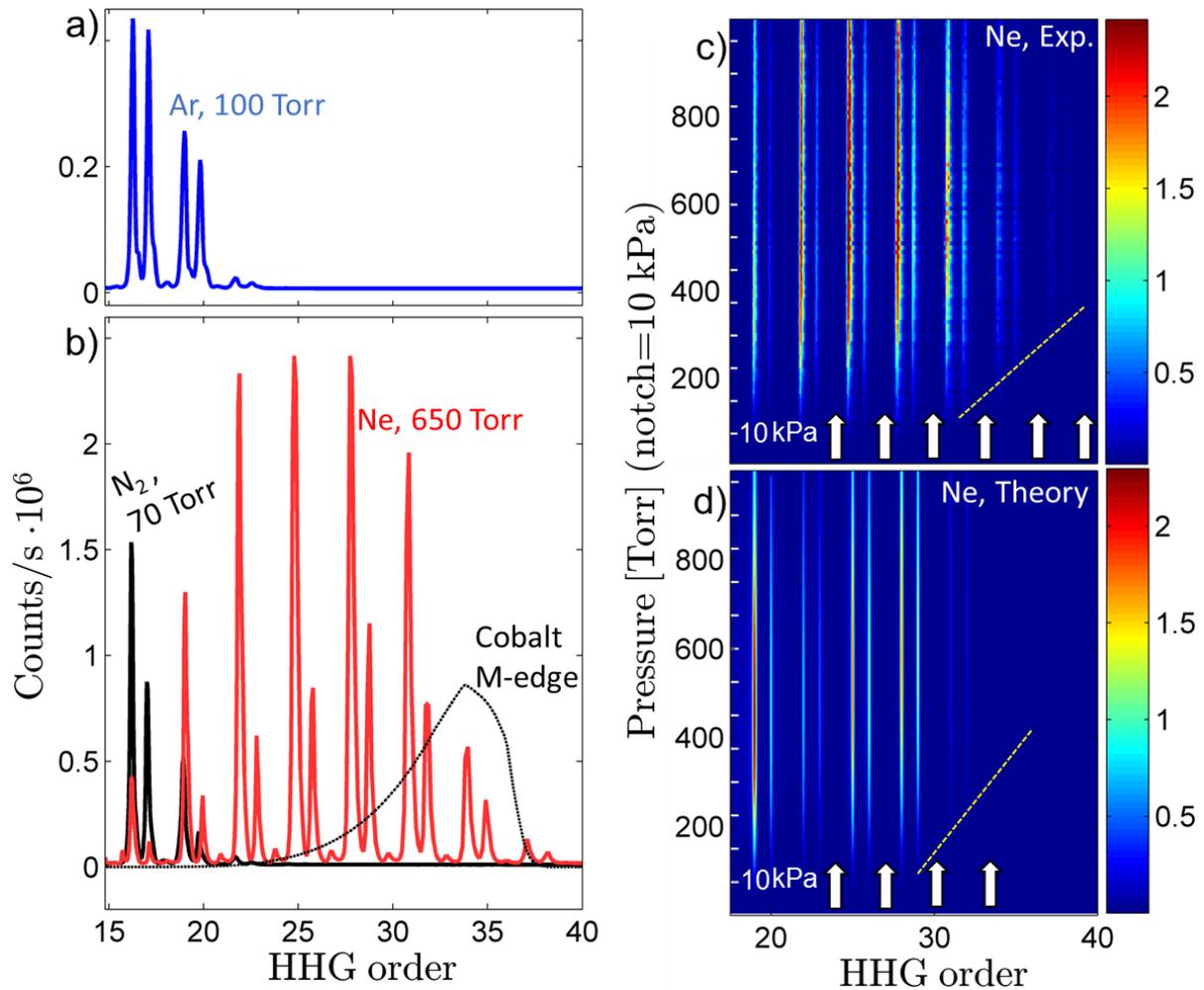

**Figure 3 | Generation of bright circularly-polarized harmonics.** (a) High harmonic spectrum from Ar using ~800 nm thick Al filter to block the laser light. (b) HHG from $N_2$ at 70 Torr (black) and Ne at 650 Torr (red solid curve) using a 400 nm thick Al filter. The HHG spectrum for Ne covers the Co *M*-edge (black dashed curve[56]). The circular polarization of the HHG is manifested by the suppression of the 3*m* harmonics (15, 18, 21…. see white arrows on (c-d). (c) Color-coded experimental and (d) simulated spectrograms as a function of the harmonic order and gas pressure for HHG with Ne. White notches are separated by 10 kPa. The color coding is in units of $10^6$ photoelectrons counts per second on the CCD. The experimental and theoretical spectrograms exhibits a known feature of phase matching in Ne: higher order harmonics are phase matched through a narrower pressure range than lower-order harmonics. (Dashed lines are a guide to the eye).

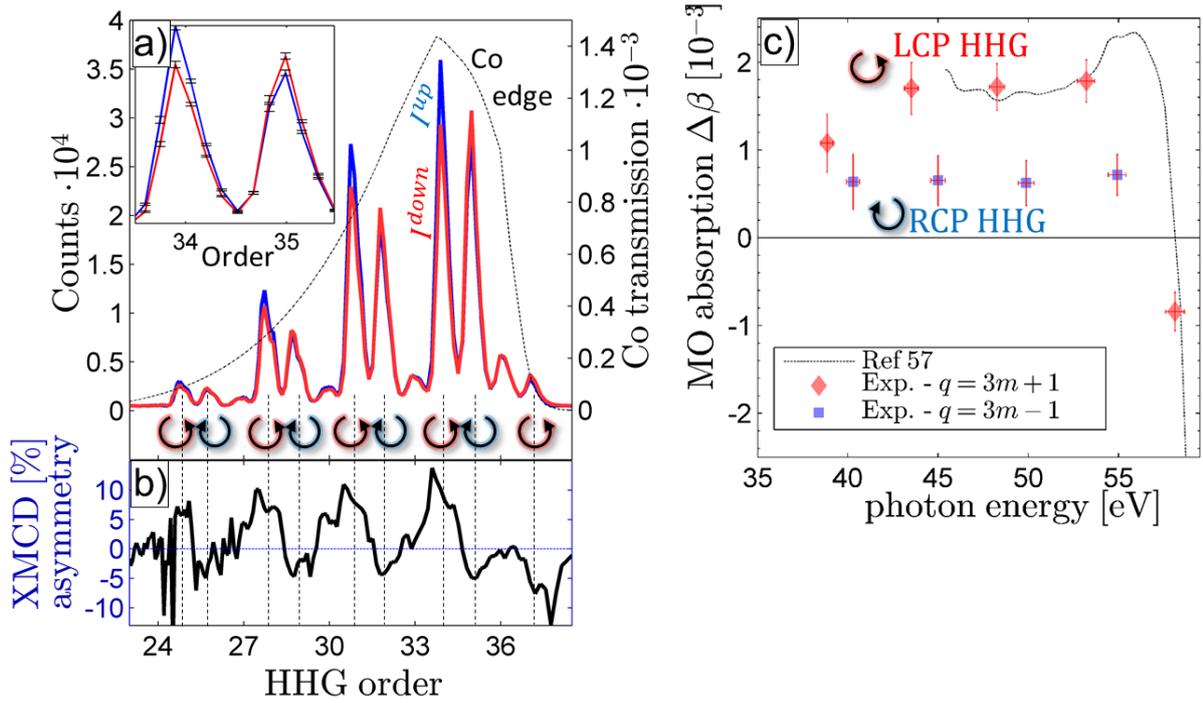

**Figure 4 | Co XMCD.** (a) Spectra of circularly polarized HHG transmitted through 100 nm of magnetized Co with the magnetization direction either "up" ($I^{up}$, blue) or "down" ($I^{down}$, red). The dashed black envelope shows the transmission spectrum of Co with units along the right-hand axis[56]. The inset shows the difference in signal between "up" and "down" magnetization for the 34th and 35th orders, including the experimental error. The helicity is indicated by circular arrows below each harmonic. (b) Experimental XMCD asymmetry defined by ($I^{up} - I^{down}$)/($I^{up} + I^{down}$). The alternating polarity of the asymmetry for adjacent harmonics is consistent with the prediction of alternating harmonic helicity. (c) The magneto-optical (MO) absorption coefficient, $\Delta\beta$, derived from XMCD asymmetry of the $3m+1$ (diamonds) and $3m-1$ (squares) harmonics. The experimental MO coefficient derived from the $3m+1$ harmonics accurately retrieves literature MO values (dashed line). Reduced intensities and XMCD asymmetry of the $3m-1$ harmonics suggests that the polarization of these harmonics is elliptical and not completely circular.

# Supplementary Information

# Generation of phase-matched circularly-polarized extreme ultraviolet high harmonics for magnetic circular dichroism spectroscopy

Ofer Kfir[1], Patrik Grychtol[2], Emrah Turgut[2], Ronny Knut[2,3], Dmitriy Zusin[2],
Dimitar Popmintchev[2], Tenio Popmintchev[2], Hans Nembach[2,3], Justin M. Shaw[3], Avner Fleischer[1,4],
Henry Kapteyn[2], Margaret Murnane[2] and Oren Cohen[1].

[1] Solid State Institute and Physics Department, Technion, Haifa 32000, Israel
[2] Department of Physics and JILA, University of Colorado and NIST, Boulder, CO 80309, USA
[3] Electromagnetics Division, National Institute of Standards and Technology, Boulder, CO 80305, USA
[4] Department of Physics and Optical Engineering, Ort Braude College, Karmiel 21982, Israel

## 1. Phase matching of circular HHG

Full phase matching of a nonlinear process corresponds to where the coherence length of the process diverges; i.e. when the right hand side of Eq. (3) equals zero. To make this possible, the effective index changes, $\Delta n_1$ and $\Delta n_2$, must have opposite sign. Furthermore, only one harmonic order can be fully phase matched for a given effective index ratio of the two colors $\Delta n_2/\Delta n_1$. In order to phase match the harmonic order $q = 3m \pm 1$, the effective index ratio should be -

$$\frac{\Delta n_2}{\Delta n_1} = -\frac{m \pm 1}{2m} \quad (S.1)$$

For example, in order to phase match the 28th or 40th HHG orders, ($m = 9, 13$) and ($\pm \to +$), the index ratio $\Delta n_2/\Delta n_1$ should be -0.55 or -0.54, respectively.

Explicit representation of the effective index change for each component of the bi-chromatic driver, $\Delta n_1$, $\Delta n_2$ is given by[59]

$$\Delta n_i = \left(\frac{\lambda_i k_i}{2\pi} - 1\right) = -\left\{\left(\frac{u_{11}^2 \lambda_i^2}{8\pi^2 a^2}\right) - P\left[(1-\eta)(n_i - 1) - \frac{1}{2\pi}\eta N_{atm} r_e \lambda_i^2\right]\right\} \quad (S.2)$$

Here, $i=1$ for the fundamental and $i=2$ for the second harmonic driver, $u_{11}$ is the modal factor[54], $a$ is the inner radius of the hollow waveguide, $\lambda_i$ and $k_i$ are the wavelength and wave-vector of $i^{th}$ field, $P$ is the gas pressure in atmospheres, $\eta$ is the ionization fraction, $r_e$ is the classical radius of the electron, $N_{atm}$ is the number density of atoms at atmospheric pressure, and $n_i$ is the index of refraction of the gas for wavelength $\lambda_i$ at atmospheric pressure. The modal term, $u_{11}^2 \lambda_i^2 / 8\pi^2 a^2$, and the plasma term $\eta N_{atm} r_e \lambda_i^2 / 2\pi$, contribute negatively to $\Delta n$, mainly to the fundamental driver, $\lambda_1$, while the normal refraction of the neutral gas, $(n_i - 1)$, contributes positively to $\Delta n$, mainly to the second harmonic field, $\lambda_2$. Since full phase matching requires that $\Delta n_2/\Delta n_1 < 0$, it is achievable when $\Delta n_1$ (long wavelength) is negative and $\Delta n_2$ (short wavelength) is positive.

The gas pressure in which high harmonics are generated acts as a continuous knob to control the HHG phase matching. At zero pressure, the effective index ratio is $\Delta n_2 / \Delta n_1 = 1/4$, where both $\Delta n_2, \Delta n_1$ are negative due to modal term, $u_{11}^2 \lambda_i^2 / 8\pi^2 a^2$. Pressurizing the hollow waveguide with a weakly ionized gas contributes positively to $\Delta n_2$, more than to $\Delta n_1$, so high enough gas pressure can make $\Delta n_2/\Delta n_1 < -0.5$, as required for phase matching of high order harmonics. Tuning the pressure, therefore, selects the HHG order to be phase matched according to Eq. (S.1) and (S.2). However, re-absorption of HHG in the gas reduces the contrast between the optimally phase matched harmonic (e.g. 28$^{th}$ order in Fig. 1b) and other harmonic orders.

Figure S.1 plots the coherence length vs. the absorption length of few HHG orders in Ne for a 1% ionization level. When the ratio of coherence length, $L_c$, to absorption length, $L_a$, is above 6 (see dashed line), the HHG flux is 90% of the fully phase matched case, and the emission is effectively phase matched. The low order HHG (19$^{th}$ and 20$^{th}$) increase at pressures around 100 Torr and are phase matched up to far above 1000 Torr, whereas the high order harmonics (40$^{th}$ and 41$^{st}$) are phase matched in a narrower pressure range. The pressure dependent spectrograms in Fig. 3c (measured) and Fig. 3d (theory) agree with this description, although the calculation takes into account additional effects, such as the HHG source density (i.e. gas density) and parasitic re-absorption in residual gas downstream from the phase matching region (details below).

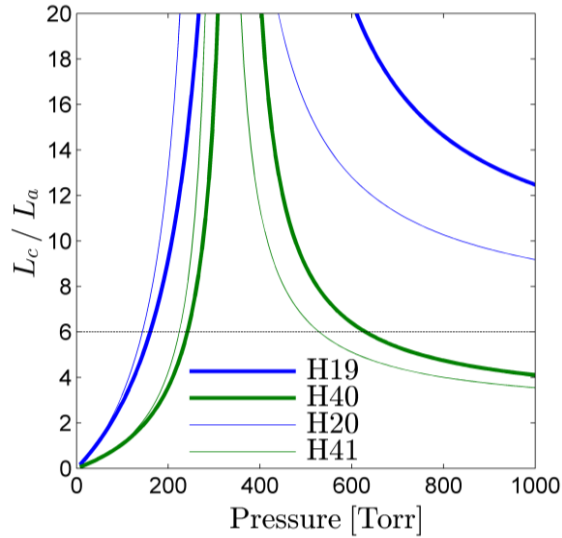

**Figure S.1** | (a) The coherence length vs. absorption length of circularly polarized harmonics in a Ne filled hollow waveguide. At $L_c/L_a = 6$ (see dashed black line), the output HHG flux is 90% of the flux under perfect phase matching condition, so the HHG are said to be phase matched. The low order HHG orders (say 19$^{th}$ and 20$^{th}$) are phase matched from about 100 Torr to above 1000 Torr, whereas the high order harmonics (40$^{th}$ and 41$^{st}$) are phase matched in a narrower pressure range. The solid and fine lines correspond to HHG of the groups $q = 3m + 1$ and $q = 3m - 1$, respectively. $L_c$ of the 19$^{th}$ order deviates from the 20$^{th}$ due to the helicity selective factor, $\pm \Delta n$, in Eq. (3).

## 2. Critical ionization for full phase matching of circularly polarized HHG

The critical ionization, $\eta_{cr}$, is the ionization fraction above which pressure tuning cannot bring about phase matching of circularly polarized HHG. Specifically, pressure tuning cannot reduce the ratio $\Delta n_2/\Delta n_1$ below -0.5. The value of $\eta_{cr}$ can be derived from Eq. (S.2) -

$$\eta_{cr} = [2(n_2 - 1) + (n_1 - 1)] \cdot \left\{2(n_2 - 1) + (n_1 - 1) + \frac{3}{4\pi} N_{atm} r_e \lambda_1^2\right\}^{-1} \quad (S.3)$$

This critical ionization differs from critical ionization of HHG driven with single laser field[54]. For HHG driven by one field in a large focus geometry, the critical ionization level is independent of the pressure, and occurs when the dispersion of the neutral atoms and plasma balance each other. (In a waveguide, the value of the critical ionization is reduced slightly below that of a large focal geometry, due to the presence of waveguide dispersion). In contrast, circularly polarized HHG driven by bi-chromatic fields requires that $\Delta n_i$ is nonzero: $\Delta n_1$ should be negative and $\Delta n_2$ positive. Since optimal $\Delta n_1$ is negative and $\Delta n_2$ is positive, $\eta_{cr}$ is larger for bi-chromatic circular driver, than for HHG driven with the fundamental only, but smaller then HHG driven by the second harmonic only.

## 3. Numerical procedure for a pressure dependent spectrogram

Simulating the single atom response to the driving bi-chromatic field is the principal step for calculating pressure dependent HHG emission. We approximate the atom as a binding potential and a single, spin-less electron, and solve the Schrödinger equation in three spatial dimensions (3D) and time. The binding potential is an inverted Gaussian given by $V_{atom}(\vec{r}) = -V_0 e^{-s|\vec{r}|^2}$, where $V_0 = 2$ and $s = 1/2$ atomic units (at.u.). The single electron initially occupies the lowest state, having ionization potential of a Ne atom, 0.794 at.u. (21.6 eV). The driving bi-chromatic field, $\vec{E}_{BC}(t)$, induces a linear potential, within the dipole approximation, so overall the potential is $V(\vec{r}, t) = -\vec{E}_{BC}(t) \cdot \vec{r} + V_{atom}(\vec{r})$. Here $\vec{E}_{BC}(t) = \vec{E}_1(t) + \vec{E}_2(t)$, $\vec{E}_1(t) = E_{1,0} e^{-(t/t_p)^6}(-\sin(\omega_0 t)\hat{x} + \cos(\omega_0 t)\hat{y})$ and $\vec{E}_2(t) = E_{2,0} e^{-(t/t_p)^6}(\sin(2\omega_0 t + \phi_2)\hat{x} + \cos(2\omega_0 t + \phi_2)\hat{y})$, where $E_{1,0}, E_{2,0}$ correspond to peak intensities of $1.3 \cdot 10^{18}[W/m^2]$ and $1.3 \cdot 10^{18}[W/m^2]$, respectively. $t_p$, correspond to 15 femtoseconds, $\omega_0 = 0.0577$ is the frequency of 790nm light in atomic units, $\phi_2$ is the phase of the second harmonic field relative to the fundamental. $\hat{x}, \hat{y}$ are the Cartesian coordinates unit vectors and $\vec{r}$ is the radius vector. The 6-power Gaussian envelope of the simulated bi-chromatic pulse is chosen to reduce spectral effects originating from variations of the pulse envelope. The emitted HHG field, $\vec{E}_{HHG}$, is calculated using Ehrenfest theorem, $\vec{E}_{HHG}(t) \propto \langle\psi(t)|\vec{\nabla} V_{atom}|\psi(t)\rangle$, where $\psi$ is the electronic wave function. The proportion constant is not accounted for since it applies globally to the entire radiation spectrum.

The HHG phase strongly depends on the effective index of the bi-chromatic driving field in the gas medium. To account for the phase differences between the fundamental and second harmonic fields, we calculate HHG emission from a single atom 100 times, where each teration uses a different phase between the fundamental and second harmonic drivers. The phase for the $h^{th}$ iteration is $\phi_{2,h} = 2\pi h/100$, where $h=1,2,3…100$, and the calculated HHG emission is assigned to an atoms positioned at $z_h = \phi_{2,h} c[2\omega_0(\Delta n_1 - \Delta n_2)]^{-1}$. The effect of the fundamental driver phase velocity is accounted for once the single atom emission, $\vec{E}_{HHG,h}(t)$, is done. Since the fundamental phase velocity in the gas filled waveguide (Eq. (S.2)) differs from free space propagation, it requires additional time, $\Delta n_1 z_h/c$, to reach the atom at $z_h$. Therefore, the emission at $z_h$ is delayed from the calculated emission, and is given by $\vec{E}_{HHG,h}(t - \Delta n_1 z_h/c)$.

The macroscopic HHG phase matched signal is a coherent sum of the single atom emissions that undergo partial re-absorption in the gas. At the end of the phase matching region, $z_{PM}$, the HHG intensity is $\left|\sum_h P\vec{E}_{HHG,h}(t - \Delta n_1 z_h/c)e^{-(z_{PM}-z_h)/L_{abs}}\right|^2$, where $L_{abs}$ is the absorption length of the HHG in the gas. The gas pressure, P, is inserted to the calculation to account for the density of HHG emitters. Upon exiting the phase matching region, the HHG field can undergo additional parasitic re-absorption in the neutral gas. This is simulated by adding absorption from a 150 $\mu m$ layer of the gas at pressure P. The pressure-dependent spectrogram in Fig. 3d is calculated by repeating the above for multiple pressure of Ne gas at an ionization fraction $\eta = 1.2\%$.

### 4. Dynamics of circularly polarized, counter rotating bi-chromatic field

The polarization of the light field is usually described by its Lissajous curve, that is, the electric field of circular, elliptical, and linear polarization follows a circle, ellipse and a line, respectively. In that sense, the polarization of a circularly polarized, counter rotating bi-chromatic field composed of a fundamental field at 790 nm and its second harmonic has a "cloverleaf" shape polarization, as described in Fig. 1a in the text. To better illustrate the dynamical aspect of this cloverleaf polarization we attach a movie showing the fundamental driver (red) second harmonic driver (blue) and the resulting field of the bi-chromatic driver (purple). (*link to a movie*)

### 5. Calculating and comparing the photon flux of circular and linear polarized harmonics.

To determine the feasibility of future XMCD pump-probe experiments using a source of circularly polarized high harmonics, we have calculated the absolute flux contained in a single circular-polarized harmonic at the Co *M*-edge below 60 eV and compared it with the flux obtained in previous ultrafast element-selective HHG experiments exploiting the transversal magneto-optical Kerr effect (T-MOKE) in reflection using linear polarized harmonics[28–32]. To this end, we have simulated the transmission as well as reflection properties of our EUV beamline components taking into account the efficiency of our CCD camera. Specifically, we have used the RAY simulation software package that is a powerful design tool developed at BESSY for synchrotron radiation beamlines[60]. Moreover, we have calibrated the efficiency of our CCD detector according to the procedure recommended by the manufacturer.

As schematically depicted in Fig. 2 of the main article, the harmonics emerging from our waveguide pass through a 200 nm thick Al filter (with a 5 nm thick Al$_2$O$_3$ layer) and the 100 nm thick Co film, which is tilted by 45° with respect to the HHG beam, before entering our spectrometer. The spectrometer consists of a toroid mirror (glass substrate coated by 100 nm of $B_4C$ and tilted at a grazing angle of 8°), a laminar gold grating patterned on top of Si substrate (with a groove density of 500 lines/mm, a groove depth of 20 nm and which is mounted at a grazing angle of 12°), two 200 nm thick Al filter (each of which having a 5 nm thick $Al_2O_3$ layer) and a CCD camera (Andor – Newton DO-920). As can been seen in Fig. S.2, this combination of components results in a theoretical throughput of $10^{-5}$-$10^{-4}$ EUV photons in the vicinity of the Co *M*-edge at 58 eV. This is comparable to the efficiency of previously reported measurements in the T-MOKE geometry.

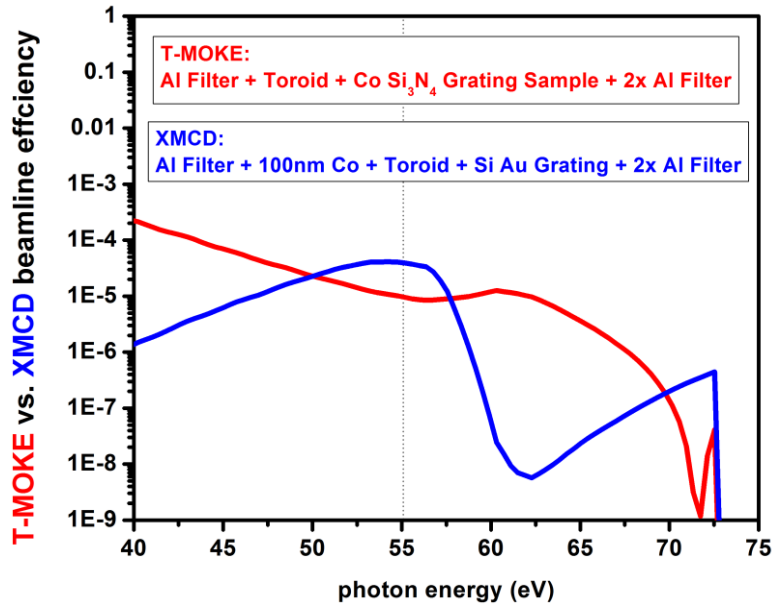

**Figure S.2 | Efficiency comparison of T-MOKE and XMCD HHG EUV beamlines**

In this spectral region, the CCD chip of the camera has a quantum efficiency of about 30%, *i.e.* about every third photon is absorbed by the back-thinned Si chip. Each absorbed photon with an energy of 55eV generates approximately 55/3.65 photoelectrons or 55/45.63 counts using the standard gain setting of our camera. Taking into account the simulated efficiency of our XMCD (T-MOKE) beamline of about $4*10^{-5}$ ($1.5*10^{-5}$) and the fact that we observe around $10^4$ counts/s in the 55 eV harmonics of both HHG setups, we obtain a photon flux of approximately $7*10^8$ photons/s in a single circular harmonic, which is only 2-3 times smaller than the flux of the very same linear harmonic in our T-MOKE experiment. Considering the very high signal-to-noise ratio that we have already attained in our static XMCD experiment as well as the relatively high effective thickness of our Co sample (140 nm) in comparison to the penetration depth of a laser excitation pulse (a few tens of nm), we conclude that ultrafast pump-probe investigations of chiral phenomena are within the capabilities of our current experimental circular HHG apparatus.

**6. Useful features of circular HHG using a counter-rotating bi-chromatic field.**
We generate bright circularly polarized harmonics by co-propagating bi-chromatic 790 nm and 395 nm (ω and 2ω) driving laser beams with counter-rotating circular polarization[50–52]. This approach has several useful properties that are beneficial for applications. First, the conversion efficiency to circularly polarized high harmonics is comparable to the conversion efficiency of traditional HHG, where linearly polarized driving lasers produce linearly polarized high harmonics (see Fig. S.2 in Ref. 50). Second, the generated spectrum consists only of circularly polarized harmonics. Third, consecutive harmonics exhibit opposite helicity. Fourth, the circular-polarization of the harmonics is manifested in the high harmonic spectrum due to angular momentum conservation rules: thus, it is not essential to directly or continuously monitor their polarization. As explained below, when the bi-chromatic driving lasers consist of the fundamental and second harmonic frequencies, absence of the $3m^{th}$ harmonics ($m$=1,2,3…) indicates that the polarizations of $3m$+1 and $3m$-1 harmonics are circular with helicity that correspond to the polarization of the fundamental and $2^{nd}$ harmonic fields, respectively. Fifth, the polarization of the generated harmonics is largely insensitive to the intensities and intensity ratio between the bi-chromatic driving lasers. This is an important feature because both the intensity of, and the intensity ratio between, the two driving fields can vary within the focal spot and along the direction of propagation. Also, this feature is important for the robustness of the method to misalignment and instability of the driving fields. Sixth, as shown below, the generation process can be fully phase matched. Finally, the mechanism for producing circularly polarized harmonics in this direct approach is based only on the circular polarization of the driving pulses and is otherwise insensitive with respect to the nonlinear medium and other properties of the driving laser beams. Thus, we expect that many experimental applications that were implemented using linearly polarized HHG can be directly transferred to circularly polarized harmonics. For example, we expect that the use of long-wavelength bi-chromatic driving lasers will extend bright circularly-polarized HHG into the keV region, which will span the water-window region as well as the magnetically sensitive *L*-shell absorption edges of the 3d ferromagnets[18].